# Proximity Effect in Nb/Au/CoFe Trilayers

Jinho Kim, Yong-Joo Doh*, and K. Char†

*School of Physics and Center for Strongly Correlated Materials Research*

*Seoul National University, Seoul, KOREA*

Hyeonjin Doh

*Department of Physics, Bk21 Physics Research Division, and Institute for Basic Science*

*Research, Sung Kyun Kwan University, Suwon 440-746, KOREA*

Han-Yong Choi

*Department of Physics, Bk21 Physics Research Division, and Institute for Basic Science*

*Research, Sung Kyun Kwan University, Suwon 440-746, KOREA, and Asia Pacific Center for*

*Theoretical Physics, Pohang 790-784, KOREA*

We have investigated the superconducting critical temperatures of Nb/Au/CoFe trilayers as a function of Au and CoFe thicknesses. Without the CoFe layer the superconducting critical temperatures of Nb/Au bilayers as a function of Au thickness follow the well-known proximity effect between a superconductor and a normal metal. The superconducting critical temperatures of Nb/Au/CoFe trilayers as a function of Au thickness exhibit a rapid initial increase in the small Au thickness region and increase slowly to a limiting value above this region, accompanied by a small oscillation of $T_c$. On the other hand, the superconducting critical temperatures of Nb/Au/CoFe trilayers as a function of CoFe thickness show non-monotonic behavior with a shallow dip feature.    We analyzed the $T_c$ behavior in terms of Usadel formalism and found that most features are consistent with the theory, although the small oscillation of $T_c$ as a function of the Au thickness cannot be accounted for.    We have also found quantitative values for the two interfaces: Nb/Au and Au/CoFe.

*Present Address: Department of NanoScience, Delft University of Technology, The Netherlands

†To whom correspondence should be addressed: kchar@phya.snu.ac.kr

**Introduction**

The proximity effect between a superconductor (S) and a normal-metal (N) has been well-known for several decades.[1,2] The superconducting transition temperature $T_c$ of SN bilayers decreases exponentially with characteristic length $\xi_N$, as the thickness $d_N$ of the overlaying N layer increases.[1,2] This is understood as an effect caused by leakage of the superconducting pairs from the S layer into the N layer. On the other hand, any proximity of a ferromagnetic layer (F) to an S layer has been found to be detrimental to superconductivity.[3-17] This has been understood as a pair-breaking effect of the magnetic moment when the superconducting pairs enter the F layer. Several theories have been developed to explain the proximity effect in SF bilayers, predicting the oscillatory $T_c$ behavior of the SF bilayer as a function of the F thickness by considering only the effect of the exchange field on the pair wavefunction without the possibility of other effects such as spin rotation or spin-flip scattering.[6,7] The characteristic length $\xi_{ex}$ of this oscillation is very small, on the order of nm because of the high exchange energy in typical ferromagnetic materials such as Co, Fe, or Ni. In order to increase this characteristic length $\xi_{ex}$, conventional magnetic materials have been alloyed with normal metals, resulting in lower Curie temperatures.[11,13,18,19]

Recently inhomogeneous superconductivity in SF multilayers has been studied experimentally and theoretically.[3,4,8,9,11-13,14] One interesting outcome is the creation of a π-junction where the phases of the two S layers sandwiching an F layer differ by π. This π-junction can be understood in the same framework as the oscillatory $T_c$ behavior of the SF bilayer. Another interesting theoretical development, the possibility of triplet superconductivity in SF multilayers, however, requires a different and new approach.[20-24] Non-collinear alignment of magnetization in an FSF system[22,23], non-uniform alignment of spin at the SF interface[21], and spin rotation or spin-flip scattering at the SF interface[20,24] have been proposed as mechanisms that can generate triplet superconductivity.

In order to study the interaction between superconductivity and ferromagnetism as we control the coupling between them, we have prepared Nb/Au/CoFe trilayers and studied $T_c$ behavior with respect to $d_{Au}$ and $d_{CoFe}$, respectively. By employing a quantitative analysis, we have found general features are consistent with the Usadel formalism, although some results cannot be understood in the framework of current proximity theory.

**Experimentals**

We prepared all our Nb/Au bilayer and Nb/Au/CoFe trilayer samples using oxidized Si wafers as substrates with a multi-source dc magnetron sputtering system at ambient temperature. The lateral size of the strip-shaped substrates was 2×7 mm². After the sputtering chamber was evacuated to 4×10⁻⁸ Torr, Nb thin films were deposited using 99.999 % pure argon gas at 6 mTorr from a solid Nb (99.95 % pure) target at a rate of 0.29 nm/sec. Following the Nb layer, the Au layer and the $Co_{60}Fe_{40}$ layers were in-situ deposited at a rate of 0.53 nm/sec and 0.17 nm/sec, respectively. For systematic variation of the CoFe layer thickness, we used the natural gradient of the sputtering rate with the stage of substrates placed in an asymmetric position relative to the center of the CoFe target. In some cases, we took samples out of the vacuum chamber after we deposited part of an Au layer in order to compare SN and SNF layers systematically by inserting half a set of samples back into the chamber for the remaining Au layer and additional F layer deposition. However, we confirmed that the brief exposure to $N_2$ gas in the loading chamber after the Au layer deposition did not alter $T_c$ of the SNF trilayers by also making the entire SNF trilayers *in-situ* and comparing the data. As a final step, all samples were covered by a 3 nm-thick Al cap layer in order to prevent possible degradation of the samples due to oxidation during measurement. All the deposition rates have been calibrated by measuring the thickness of thick films by a profilometer.

Typical rms roughnesses of the Nb layer, Nb/Au bilayer, Nb/Au/CoFe trilayers were 0.17 nm, 0.24 nm, and 0.19 nm, respectively, when measured by an atomic force microscope. The large increase in the roughness of the Nb/Au layer is probably due to the difficulty associated with Au layer wetting on the Nb surface. The uniformity of $T_c$ of a set of identical Nb films made by simultaneous deposition is within 20 mK. The superconducting transition temperature $T_c$ was measured resistively in a standard four-terminal configuration and determined from the R(T) curves using the 10 % criterion. The measurements were performed by standard d.c. or a.c. lock-in techniques, using current magnitudes of 0.1 mA.

**Results and analysis**

Near the superconducting transition temperature $T_c$, R(T) curves of Nb/Au bilayers and Nb/Au/CoFe trilayers are shown in Fig. 1, where $d_{Au}$ was varied for fixed $d_{Nb}$ = 23 nm and $d_{CoFe}$ = 10 nm. The resistance was normalized by the normal state value at T = 10 K. The transition width, corresponding to the temperature difference between 90 % and 10 % of the normalized resistance, is about 10 mK (19 mK) for Nb/Au (Nb/Au/CoFe) series.

With increasing $d_{Au}$, the $T_c$'s of the Nb/Au bilayers show a monotonous decrease and have nearly a saturated value above $d_{Au}$ = 150 nm. In Nb/Au/CoFe trilayers, however, $T_c$'s show the opposite behavior, rapid increase until $d_{Au}$ = 10 nm and gradual increase from $d_{Au}$ = 10 nm to 150 nm. For the case of $d_{Au}$ = 0 nm, $T_c$ of Nb/CoFe bilayer is 4.47 K. Both the Nb/Au bilayer and the Nb/Au/CoFe trilayer have the same $T_c$ values within the experimental error for $d_{Au}$ = 250 nm.

The contrast in $T_c$ behavior between the Nb/Au bilayer and the Nb/Au/CoFe trilayer series with varying $d_{Au}$ can be seen clearly in Fig. 2. The SN bilayer series follows the form of an exponential decay, while the SNF trilayer series exhibits very rapid increase of $T_c$ as soon as the Au layer is inserted and then approach slowly to the limiting value that is in good agreement with the limiting value of the corresponding SN series.

To analyze the $T_c$ behaviors of SN and SNF systems in a consistent manner, we used the Usadel equations.[25] The calculation procedure for $T_c$ for the SNF trilayer can be generalized from the result of the SF bilayer[7,26], which will be presented below. Here, we did not consider the triple pairing states included in Ref. 26. The SN bilayer can be considered as a limiting case in which $d_F$ is zero. The Usadel equation for a SNF trilayer near $T_c$ may be written as

$$\pi k_B T_{cS} \xi_S^2 \frac{d^2}{dx^2} f_S(x; i\omega_n) = |\omega_n| f_S(x; i\omega_n) - \Delta(x), \quad (x > 0) \quad (1)$$

$$\pi k_B T_{cS} \xi_N^2 \frac{d^2}{dx^2} f_N(x; i\omega_n) = |\omega_n| f_N(x; i\omega_n), \quad (-d_N < x < 0) \quad (2)$$

$$\pi k_B T_{cS} \xi_F^2 \frac{d^2}{dx^2} f_F(x; i\omega_n) = |\omega_n| f_F(x; i\omega_n) + i\,\mathrm{sgn}(\omega_n) E_{ex} f_F(x; i\omega_n), \quad (-d_F - d_N < x < -d_N)$$

$$(3)$$

where $f_{S(N,F)}$ is the anomalous Green's function of the Usadel equation for a superconductor (normal-metal, ferromagnet) region and $\xi_{S(N,F)}$ is the characteristic length defined from the diffusion constants of a superconductor (normal-metal, ferromagnet).

$$\xi_S = \sqrt{\frac{\hbar D_S}{2\pi k_B T_{cS}}}, \quad \xi_N = \sqrt{\frac{\hbar D_N}{2\pi k_B T_{cS}}}, \quad \text{and} \quad \xi_F = \sqrt{\frac{\hbar D_F}{2\pi k_B T_{cS}}}. \quad (4)$$

The proper boundary conditions are

$$\frac{d}{dx} f_F(-d_F - d_N) = \frac{d}{dx} f_S(d_S) = 0, \quad (5)$$

$$\xi_N \frac{d}{dx} f_N(-d_N) - \gamma_{NF} \xi_F \frac{d}{dx} f_F(-d_N) = 0, \quad (6)$$

$$\xi_S \frac{d}{dx} f_S(0) - \gamma_{SN} \xi_N \frac{d}{dx} f_N(0) = 0, \tag{7}$$

$$f_N(-d_N) - f_F(-d_N) = \gamma_b^{NF} \xi_F \frac{d}{dx} f_F(-d_N), \tag{8}$$

$$f_S(0) - f_N(0) = \gamma_b^{SN} \xi_N \frac{d}{dx} f_N(0), \tag{9}$$

where

$$\gamma_{NF} \equiv \frac{\rho_N \xi_N}{\rho_F \xi_F}, \quad \gamma_{SN} \equiv \frac{\rho_S \xi_S}{\rho_N \xi_N}, \quad \gamma_b^{NF} \equiv \frac{R_b^{NF} A}{\rho_F \xi_F}, \quad \gamma_b^{SN} \equiv \frac{R_b^{SN} A}{\rho_N \xi_N}. \tag{10}$$

We calculate the $T_c$ from the self-consistency relation,

$$\ln \frac{T_{cS}}{T} \Delta(x) = \pi k_B T \sum_{\omega_n} \left[ \frac{\Delta(x)}{|\omega_n|} - f_S(x; i\omega_n) \right]. \tag{11}$$

The procedure for solving the equations is the same as that used in Ref. 7. We solve the Usadel equation for $-d_F - d_N < x < 0$ with the boundary conditions. This results in the following relation

$$\xi_S \frac{d}{dx} f_S^{(+)}(0; i\omega_n) = W(i\omega_n) f_S^{(+)}(0; i\omega_n), \tag{12}$$

where

$$f_{S(N,F)}^{(\pm)}(x; \omega_n) = f_{S(N,F)}(x; \omega_n) \pm f_{S(N,F)}(x; -\omega_n) \tag{13}$$

and

$$W(i\omega_n) = \gamma_{SN} \frac{A_S(\gamma_b^{SN} + \mathrm{Re}\, B_{SN}) + \gamma_{SN}}{A_S \left| \gamma_b^{SN} + B_{SN} \right|^2 + \gamma_{SN}(\gamma_b^{SN} + \mathrm{Re}\, B_{SN})}. \tag{14}$$

The $A_S$ and $B_{SN}$ are defined as

$$A_S(i\omega_n) = k_S \xi_S \tanh k_S d_S \tag{15}$$

$$B_{SN}(i\omega_n) = \left[k_N \xi_N \tanh k_N (d_N + x_0)\right]^{-1} \tag{16}$$

$$\tanh k_N x_0 = \frac{1}{k_N \xi_N} \frac{\gamma_{NF}}{\gamma_b^{NF} + B_F} \tag{17}$$

$$B_F(i\omega_n) = \left[k_F \xi_F \tanh k_F d_F\right]^{-1}. \tag{18}$$

Here, $k_{S(N,F)}$ is the wave number in the superconductor (normal-metal, ferromagnet) defined in the following way.

$$k_S = \frac{1}{\xi_S}\sqrt{\frac{|\omega_n|}{\pi k_B T_{cS}}}, \quad k_N = \frac{1}{\xi_N}\sqrt{\frac{|\omega_n|}{\pi k_B T_{cS}}}, \quad \text{and} \quad k_F = \frac{1}{\xi_F}\sqrt{\frac{|\omega_n| + iE_{ex}\,\text{sgn}\,\omega_n}{\pi k_B T_{cS}}}. \tag{19}$$

We use the fundamental solution (Green's function) for the inhomogeneous equation (1). The $T_c$ is obtained by the largest T of the following equation,

$$\Delta(x)\ln\frac{T_{cS}}{T} = 2\pi k_B T \sum_{\omega_n > 0} \int_0^{d_S} dy \left(\frac{\delta(x-y)}{|\omega_n|} - G(x,y;i\omega_n)\right)\Delta(y) \tag{20}$$

where $G(x, y; i\omega_n)$ is the Green's function of the inhomogeneous equation written as

$$G(x,y;i\omega_n) = \frac{k_S/|\omega_n|}{\sinh k_S d_S + (W(i\omega_n)/k_S \xi_S)\cosh k_S d_S} \times \begin{cases} v_1(x)v_2(y), (0 < x < y < d_S) \\ v_1(y)v_2(x), (0 < y < x < d_S) \end{cases}, \tag{21}$$

with

$$v_1(x) = \cosh k_S x + \frac{W(i\omega_n)}{k_S \xi_S}\sinh k_S x, \quad \text{and} \tag{22}$$

$$v_2(x) = \cosh k_S (x - d_S). \tag{23}$$

The above integral equation (20) can be transformed into a simple eigenvalue problem by discretizing the integration into the summation. Then, we can obtain the $T_c$ by numerical calculation.

For the calculation of the $T_c$ of Nb/Au bilayers, the parameters to be determined are the resistivity $\rho$ and the dirty limit coherence length $\xi$ of each layer, the $T_c$ of the Nb single layer $T_{cS}$, and the parameter representing the interface between the S and N layers $\gamma_b^{SN}$. We measured $\rho_{Nb} = 15.2\,\mu\Omega\text{cm}$, $\rho_{Au} = 2.3\,\mu\Omega\text{cm}$ and $T_{cS}$ = 7.73 K from separate experiments. The resistivities are the residual resistivity values measured at T = 10 K. The remaining parameters were determined from the process finding a calculation result which fits our $T_c$ data for Nb/Au bilayers. The best result for the calculation was obtained with the parameters $\xi_{Nb} \approx 7.0\,\text{nm}$, $\xi_{Au} \approx 85\,\text{nm}$, and $\gamma_b^{SN} \approx 1.15$. This result is represented in Fig. 2 as a solid line. As can be seen in Fig. 2, the theoretical calculation agrees well with our data.

The mean free path of the Nb inferred from the coherence length, obtained above by substituting $v_F = 0.56 \times 10^6\,\text{m/sec}$[27] for the Fermi velocity, is $l_{Nb} \approx 1.7\,\text{nm}$. This indicates that Nb in this experiment satisfies the dirty limit condition. On the other hand, the mean free path of Au is estimated to be $l_{Au} \approx 98\,\text{nm}$, when substituting $v_F = 1.4 \times 10^6\,\text{m/sec}$, the Fermi velocity value obtained from a free electron model.[28] This mean free path is slightly larger than the coherence length of Au. Thus, we cannot exclude the possibility that the dirty limit assumption is not appropriate in this calculation. The interface parameter $\gamma_b^{SN} \approx 1.15$ is a somewhat large value considering that this parameter represents the ratio of the resistance of the interface itself to the resistance in the normal metal felt by the Cooper pairs. This large interface resistance is probably due to intermixing of Nb and Au at the atomic level. When Al or Cu was used instead of Au, the interface parameters were much smaller.[29,30]

The $T_c$ of Nb/Au/CoFe trilayers was analyzed using the method we described. In this method, the behavior of the order parameter in all three layers was calculated taking the effect of the two interfaces into account. From this, the $T_c$ of the SNF trilayer could be obtained. Using this method, we calculated the $T_c$ of Nb/Au/CoFe trilayers as a function of $d_{Au}$. For the Nb and Au layers and SN interface, the same parameters as were determined from the calculation of $T_c$ for the SN bilayers mentioned above were used. For a CoFe layer, we used the values for the resistivity $\rho_{CoFe} = 14.8\ \mu\Omega cm$, and a coherence length of the CoFe layer $\xi_{CoFe} \approx 14.4$ nm, and the Curie temperature for CoFe as 1152 K obtained from the fit of the $T_c$ behavior of Nb/CoFe bilayers in our previous report.[14] Notice that the definition of $\xi_{CoFe} = \sqrt{\hbar D_{CoFe}/2\pi k_B T_{cS}}$ is different from the characteristic length scale of the modulation of the order parameter in the F layer $\xi_{ex} = \sqrt{\hbar v_F l_{CoFe}/\pi E_{ex}}$. The interface parameter between the N and F layers $\gamma_b^{NF}$ was determined from the fitting of the $T_c$ data of a Nb/Au/CoFe trilayer to the theory. The best calculation result was obtained, yielding $\gamma_b^{NF} = 0.5$. This result is depicted by a dashed line in Fig. 2.

As can be seen in Fig. 2, the calculation result exhibits much higher $T_c$ values than the data when $d_{Au} = 0$. The reason for this discrepancy is as listed below. In the calculation of $T_c$ for the SNF trilayer, we take two interfaces into account. They affect the $T_c$ of the system even though the thickness of the normal metal layer goes to zero in the calculation. But the sample without the normal metal layer contains only one interface between the S and F layers. The interface parameter of the SF boundary obtained from the fitting of the Nb/CoFe bilayer is $\gamma_b^{SF} = 0.34$ [14], which is smaller than the values $\gamma_b^{SN}$ for Nb/Au and $\gamma_b^{NF}$ for Au/CoFe. Therefore, the calculated $T_c$ value at $d_{Au} = 0$ exhibits higher value than the data.

In this calculation, the effect of the interface is implied in the boundary condition of the Usadel equations, which can be described by interface resistance. Therefore, the microscopic structure of the interface is ignored in this formulation. In real situation, we need some distance for formation of the metal-metal interface. However, this distance cannot be considered in the calculation. Therefore, the jump in $T_c$ is unavoidable as soon as the effect of the two interfaces starts to be included in the calculation.

The magnified view of the rapid increase of $T_c$ in Nb/Au/CoFe trilayer is seen in the inset of Fig. 2. In a very small length scale of the Au layer, the $T_c$ shows a rapid but monotonous increase with increasing $d_{Au}$ and have a saturated value above $d_{Au} = 3$ nm. Because this increase cannot be understood using the method used above, we analyzed the $T_c$ behavior in this region in a different way. As an analogy with the de Gennes-Werthamer theory of SN bilayers, we adopted a qualitative fit using $T_{cSNF} = T_{\lim} + C^* \exp(-2d_N/\xi^*)$ in which the fitting parameter $C^*$ has a negative sign with $O(1)$. The fit in the inset of Fig. 2 of the $T_c$'s in SNF trilayers displays exponential behavior with a characteristic length of $\xi^* \approx 2$ nm for both Nb/Au/CoFe trilayer series. This initial rapid increase may be an outcome of the establishment of the two interface layers, although the length of 2nm seems too large a value for formation of a metal-metal interface.

After this rapid increase of $T_c$ until the Au layer thickness reaches 5 nm, the $T_c$ values start to deviate from the exponential relation it followed. In Fig. 3 the solid line represents the exponential fit with the characteristic length of $\xi^* \approx 2$ nm that we found for the inset of Fig. 2, while the dashed line is the result of the calculation mentioned above with a completely different characteristic length scale, $\xi = 85$ nm. After about $d_{Au}$=10 nm, the exponential approach to a limiting value seem to have switched to another form with a much longer characteristic length scale, which is the normal coherence length of Au.

In addition to this switching to a new exponential form, the $T_c$ values seem to go through oscillations as a function of Au layer thickness in its range from 20 nm to 110 nm. The inset of Fig. 3 illustrates this point more clearly, for which we have fabricated and measured three series of trilayer samples in the range of 20 nm < $d_{Au}$ < 110 nm. $\Delta T_c$ on the vertical axis means the difference in $T_c$ relative to the value at $d_{Au}$ = 110 nm. On top of a gradual increase, there are clearly oscillations in all three series with an oscillation period of about 21.6 nm. Recognizing the lack of sufficient data for this length scale, we have repeated this experiment with much denser data in the range of 20 nm < $d_{Au}$ < 60 nm on one set of series and confirmed the same oscillation period. This $T_c$ oscillation as a function of $d_{Au}$ does not depend on the sample preparation methods and its amplitude is larger than the experimental error bar in our experiment, though its amplitude is dependent on $d_{Nb}$. Trilayers with thinner Nb layers exhibit a larger amplitude, suggesting that it is unlikely that this is an artifact of fabrication or measurement technique. This oscillation of $T_c$ as a function of N layer thickness may suggest that the N layer acts as a weak ferromagnetic material, somewhat surprising given the large thickness of the N layer, several tens of nm. Alternatively it may not have anything to do with the magnetism. It may represent some electronic interference effect inside the N layer, although the length scale of 20 nm seems too small considering the Fermi velocity. We have not observed this $T_c$ oscillation when Al or Cu was used instead of Au.[29,30] This oscillatory behavior can by no means be understood in the framework of the conventional theory.

In Fig. 4 $T_c$'s of Nb/Au/CoFe trilayer systems are shown as a function of $d_{CoFe}$ with $d_{Nb}$ = 24 nm and $d_{Au}$ = 5, 10, and 30. There are small plateaus of $T_c$ near $d_{CoFe}$ = 0.5 ~ 1.2 nm but, regardless of the Au layer thickness, the $T_c$'s become a minimum around $d_{CoFe}$ = 3 nm and then eventually approach a limiting value. The plateaus may be attributed to the increase of rms roughness due to the lack of good wetting of the Au layer and the resulting magnetic dead layer of CoFe with the order of roughness.[10]  However, the $d_{CoFe}$ value around 3 nm where $T_c$'s show minimums for all thicknesses of Au layer, qualitatively consistent with the feature in Nb/CoFe bilayers without the Au layer which can be found in our previous report[14], suggests that the FFLO framework[31,32] remains valid in the Nb/Au/CoFe trilayer systems, although the magnitude of the effect starts to decrease as the thickness of the N layer increases.

This can be seen in Fig. 5, where the $T_c$'s of Nb/Au/CoFe trilayer systems are shown as a function of $d_{CoFe}$ with $d_{Nb}$ = 24 nm $d_{Au}$ = 10 nm with a calculation result with the same parameters as in the calculation in Fig. 2 except for the slightly different $T_{cS}$ = 7.86 K.  All the interface parameters are also consistent with the values we obtained for Fig. 2.

In order to make sure that all the parameters including the interfaces are consistent, we have repeated our experiment with thinner Nb.  In Fig. 6, we present $T_c$'s of Nb/Au/CoFe trilayer systems as a function of $d_{CoFe}$ with $d_{Nb}$ = 15 nm and $d_{Au}$ = 10 and 50 nm. The lines are the calculation results with the same parameters as in the calculation in Fig. 2 except for different $T_{cS}$ = 7.35 K and 7.73 K, respectively, due to the different thicknesses of the initial Nb layer. Again the general features can be fitted well except for the very thin CoFe region, probably due to the weakened magnetism as mentioned above for Fig. 4.

In summary, we have studied the superconducting critical temperatures of Nb/Au/CoFe trilayer systems as a function of Au and CoFe layer thickness in order to control the coupling between the superconductivity and the ferromagnetism. When analyzed with a theory based on the conventional framework[26], the general $T_c$ behavior of Nb/Au bilayers was in good agreement and we were able to find materials and interface parameters that are consistent throughout our experiments. However, there are some unexpected aspects that could not be explained in the framework of the conventional proximity theory in the $T_c$ behavior of Nb/Au/CoFe trilayers; there is a small oscillation of $T_c$ as a function of $d_{Au}$ with a period of about 20 nm. In addition, the length scale of $d_{Au}$ for the initial rapid increase of $T_c$ seems too large to be regarded as the length necessary for the formation of a metal-metal interface.

This work is partially supported by KOSEF through CSCMR and by MOST through the Tera-level Nano Frontier Program.

**Fig. 1**

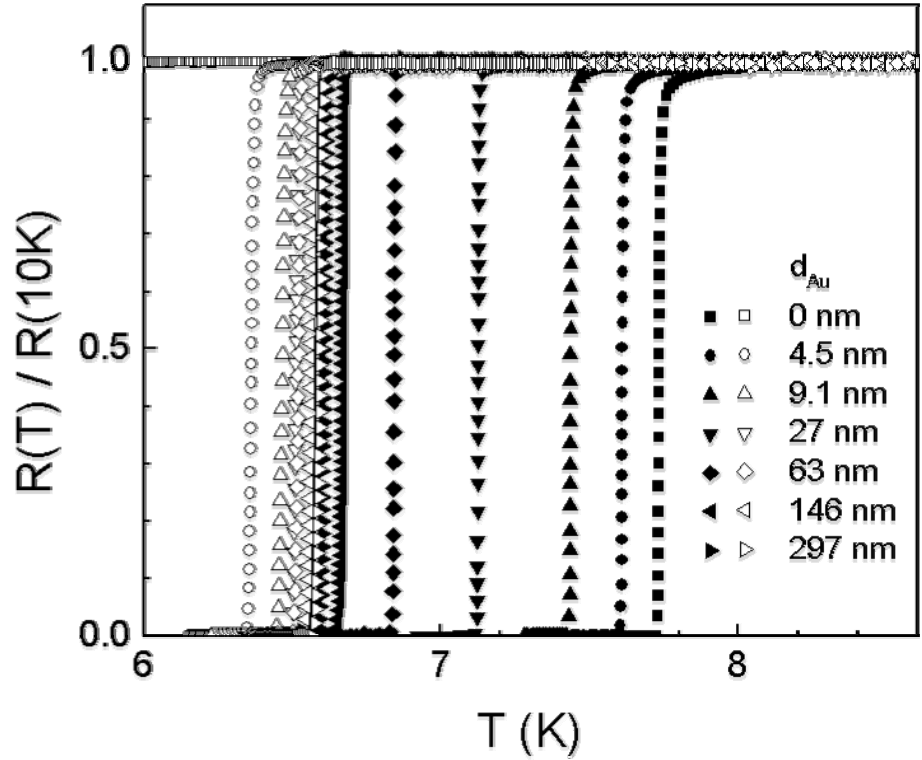

Fig. 1. Normalized R(T) curves of Nb/Au (solid symbols) bilayer and Nb/Au/CoFe (open symbols) trilayer samples near $T_c$ with varying $d_{Au}$ for $d_{Nb}$ = 23 nm and $d_{CoFe}$ = 10 nm. The resistance was normalized by the value in a normal state at T = 10 K. We determined $T_c$ using the 10 % criteria, explained in the text. For Nb/CoFe bilayer (open rectangle), $T_c$ is about 4.47 K (not shown here).

**Fig. 2**

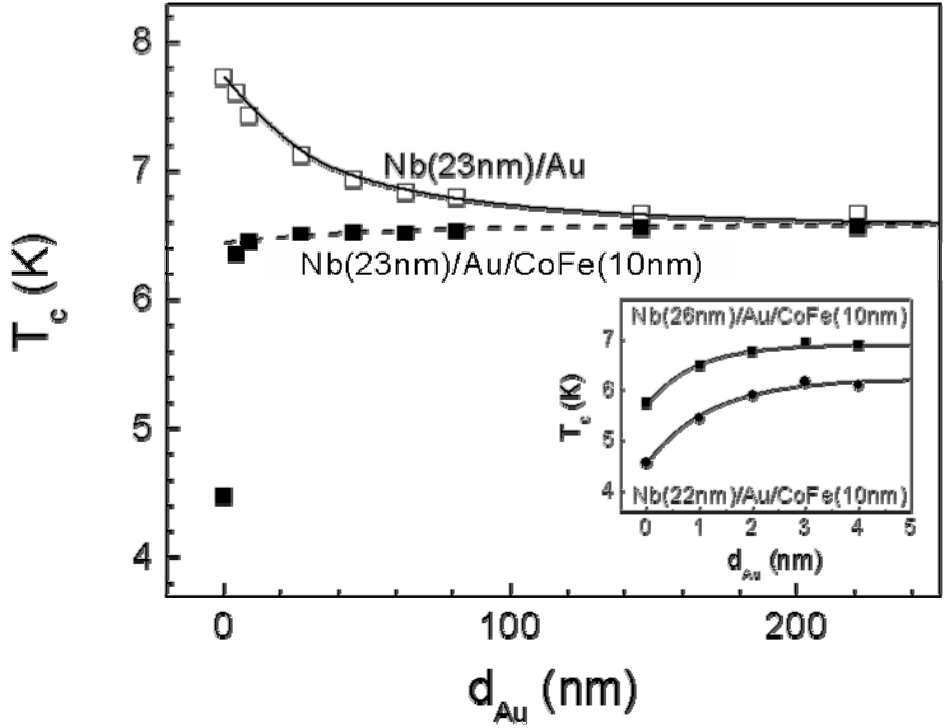

Fig. 2. Au insertion layer's thickness dependence of the superconducting transition temperature $T_c$ of Nb/Au bilayers (empty symbol) and Nb/Au/CoFe trilayers (solid symbol) with $d_{CoFe} = 10$ nm for $d_{Nb} = 23$ nm. The solid(dashed) line is a result of the calculation for the Nb/Au(Nb/Au/CoFe) system. Inset: Magnified view of $T_c(d_{Au})$ of Nb/Au/CoFe trilayers for $d_{Nb} = 22$ nm (solid circle) and 26 nm (solid square), respectively. The solid lines are the results of a fit to a first order exponential decay with a characteristic length of $\xi^* \approx 2$ nm.

**Fig. 3**

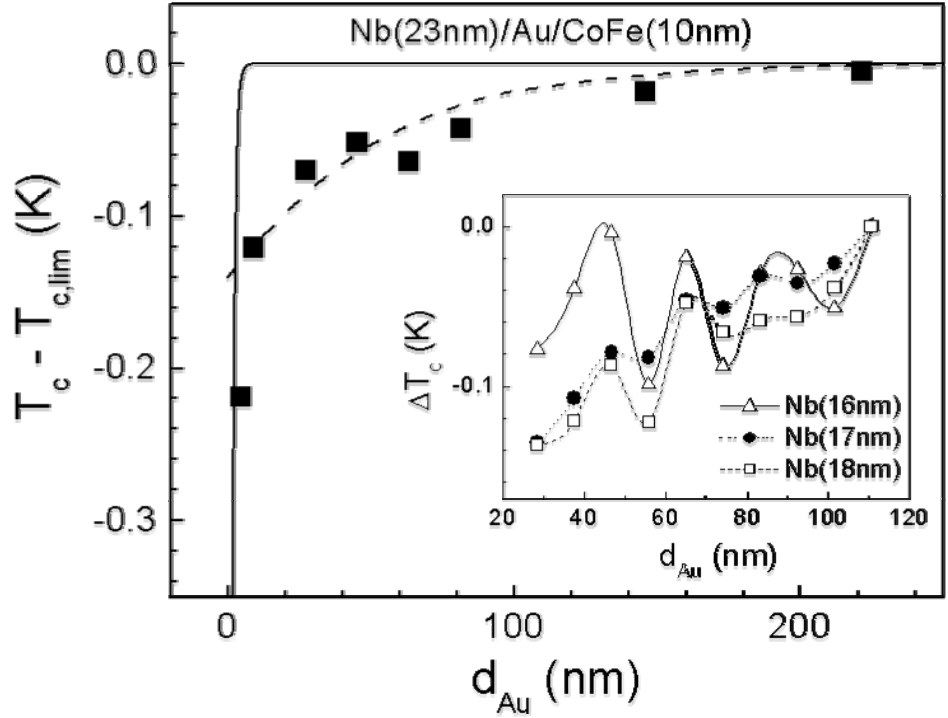

Fig. 3.  Rescaled view of $T_c(d_{Au})$ of Nb/Au/CoFe trilayers for $d_{Nb}$ = 23 nm (solid square) in Fig. 2. The solid line is a fitted result with $\xi^* \approx$ 2 nm and the dashed line is the result of a calculation with $\xi^* \approx$ 85 nm. Inset: $T_c$ variation of SNF trilayers with respect to $d_{Au}$ with varying $d_{Nb}$ = 16, 17, and 18 nm. $\Delta T_c$ means the difference of $T_c(d_{Au})$ relative to $T_c(d_{Au}$ = 110 nm) for the series.

**Fig. 4**

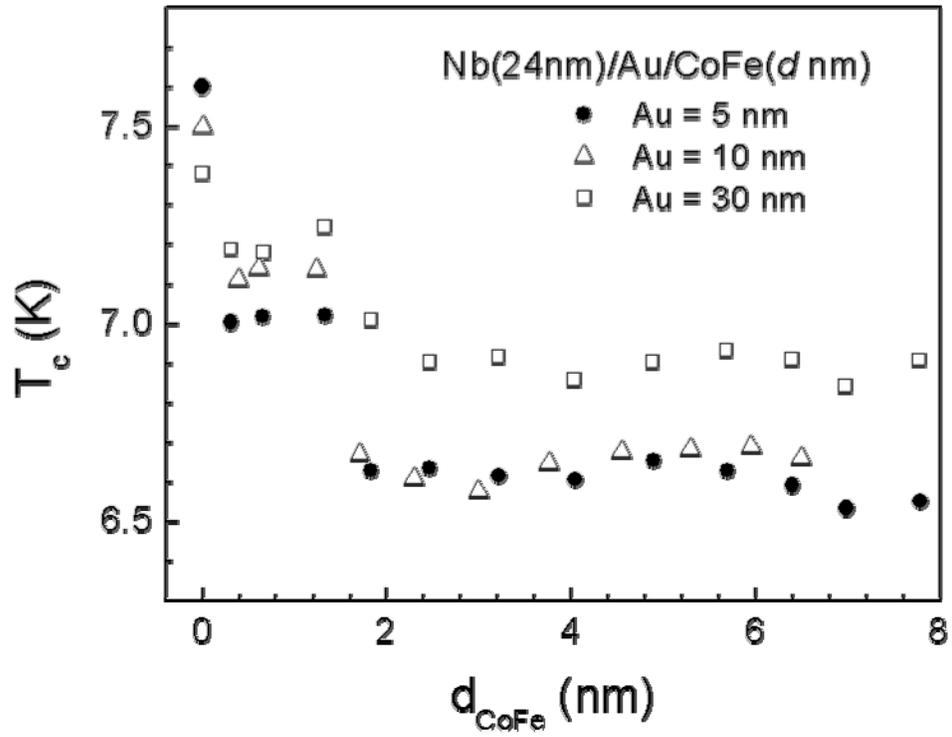

Fig. 4. $T_c$ behavior of Nb/Au/CoFe trilayers with respect to $d_{CoFe}$ with $d_{Au}$ = 5, 10, and 30 nm respectively for $d_{Nb}$ = 24 nm.

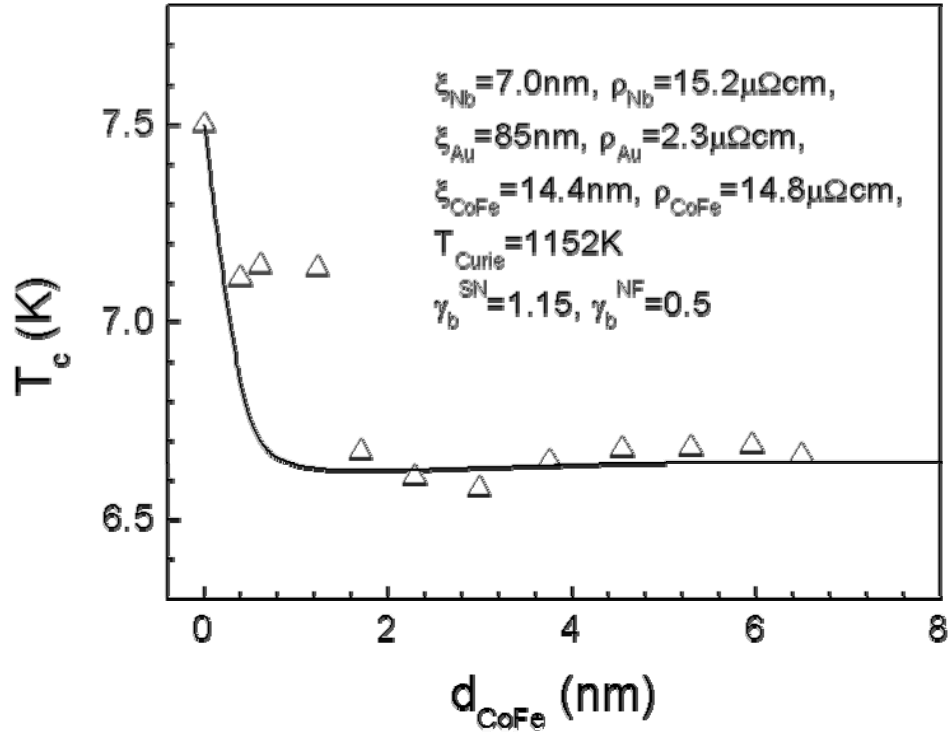

Fig. 5. $T_c$ behavior of Nb/Au/CoFe trilayers with respect to $d_{CoFe}$ for $d_{Au}$ = 10 nm and $d_{Nb}$ = 24 nm. The solid line is the calculation result explained in the text.

**Fig. 6**

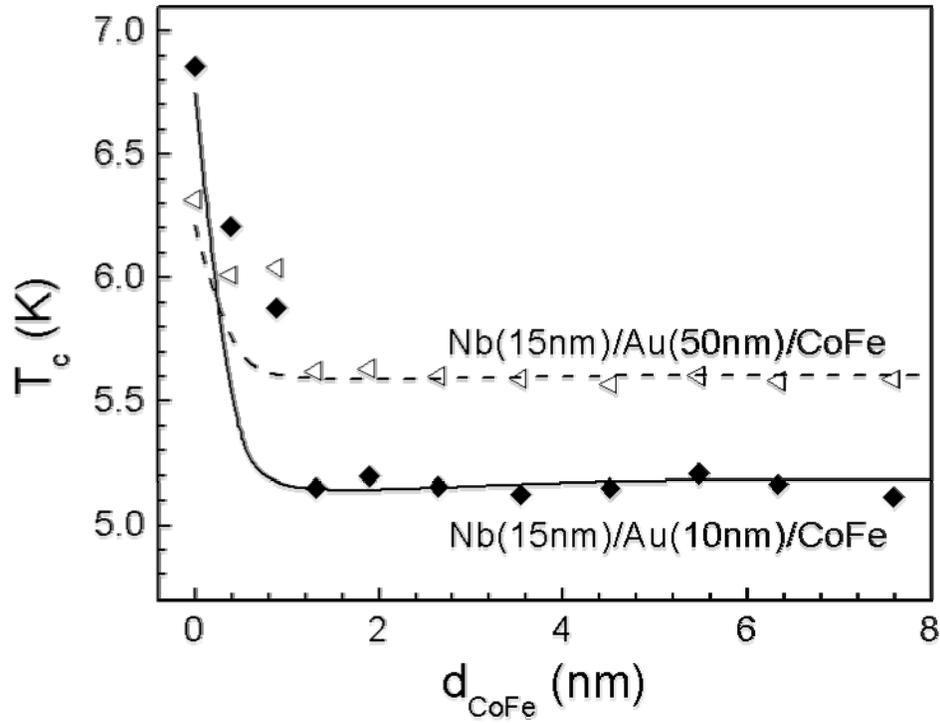

Fig. 6. $T_c$ behavior of Nb/Au/CoFe trilayers with respect to $d_{CoFe}$ with fixed $d_{Nb}$ = 15 nm for $d_{Au}$ = 10 and 50 nm , respectively. The lines are calculation results.